\documentclass[twocolumn,showpacs,preprintnumbers,amsmath,amssymb,prb,superscriptaddress]{revtex4}
\usepackage{graphicx,curves,epic}
\preprint{Submitted to Physical Review B}

\newcommand{\ba}{{\bf a}}

\newcommand{\bq}{{\bf q}} 
\newcommand{\bv}{{\bf v}} 
\newcommand{\bp}{{\bf p}} 
\newcommand{\br}{{\bf r}}

\newcommand{\bu}{{\bf u}}

\newcommand{\bR}{{\bf R}}

\newcommand{\cL}{{\mathcal L}}

\newcommand{\cP}{{\mathcal P}}

\begin{document} 
\title{Is weak temperature dependence of electron dephasing possible?} 
\author{V. V. Afonin}
\affiliation{A. F. Ioffe  Physico-Technical Institute of Russian
  Academy of Sciences, 194021
  St. Petersburg, Russia}
\author{J. Bergli} 
\email{joakim.bergli@fys.uio.no}
\affiliation{Department of Physics, University of Oslo, PO Box 1048
  Blindern, 0316 Oslo, Norway}
\author{Y. M. Galperin}
\affiliation{Department of Physics, University of Oslo, PO Box 1048
  Blindern, 0316 Oslo, Norway}
\affiliation{A. F. Ioffe  Physico-Technical Institute
of Russian
  Academy of Sciences, 194021
  St. Petersburg, Russia}
\author{V. L. Gurevich}
\author{V. I. Kozub}
\affiliation{A. F. Ioffe  Physico-Technical Institute
of Russian
  Academy of Sciences, 194021
  St. Petersburg, Russia}

 
\date{\today} 
 
\begin{abstract} 
The first-principle theory of electron dephasing by 
disorder-induced two state fluctuators is developed. 
There exist two mechanisms of dephasing. First,
dephasing  occurs due to direct transitions between the defect
levels caused  by inelastic electron-defect scattering. 
The second mechanism is due to violation of the time reversal symmetry
caused by time-dependent fluctuations of the scattering potential.
These fluctuations originate from an interaction between the dynamic
defects and conduction electrons forming a thermal bath. 
The first contribution to the dephasing rate saturates as temperature 
decreases. The second contribution does not saturate, although its 
temperature dependence is rather weak, $\propto T^{1/3}$. 
The quantitative 
estimates based on the experimental data  show that these mechanisms
considered can explain the weak temperature dependence of the dephasing
rate in some temperature interval. However, below some temperature
dependent on the model of dynamic defects the dephasing rate tends
rapidly to zero. The relation to earlier studies of the 
dephasing caused by the dynamical defects is discussed. 
\end{abstract} 
 
\pacs{73.20.Dx} 
\maketitle
 
\section{Introduction}  \label{introduction} 
The problem of dephasing of electron states in low-dimensional 
structures is in focus of interests of many research groups. This is 
due to novel experiments on the Aharonov-Bohm effect in specially 
designed mesoscopic circuits\cite{heiblum1,heiblum2}  
and on weak localization 
magnetoresistance in low-dimensional samples,\cite{mw} as well as to 
new theoretical discussions of dephasing.\cite{gz,aag,mis,ifs,RDZ,Ahn} In 
particular, dephasing due to defects with internal degrees of freedom 
as a source of dephasing were recently addressed.\cite{mis,ifs} 
According to the model discussed in Ref.~\onlinecite{ifs}, a 
temperature interval can exist in which the dephasing rate is almost 
temperature-independent.  
 
In this work we revisit the  dephasing due to dynamic defects which 
interact with electrons and tunnel between their two states due 
to interaction with some thermal bath. Examples of such defects 
are disorder-induced two-state fluctuators\cite{ahvp,Black} present in 
any disordered material, impurities with a non-compensated spin, 
etc. These defects produce a random time-dependent field and in this way 
they violate the time-reversal symmetry of the problem. According to a 
conventional opinion, this property is sufficient to produce dephasing. 
However, this is true only under 
the condition that a typical defect relaxation time 
is shorter that the time during which the electron interference 
pattern is formed. Indeed, if the defects do not change their state 
during the pattern formation they act as static ones and can contribute 
to the interference only in a constructive way.\cite{ifs} 
 
The purpose of this paper is to develop a systematic theory of weak 
localization with dephasing due to dynamic defects interacting with 
electrons  
which results in a smooth temperature dependence at relatively 
low temperatures $T$. The dynamic defects are specified as two-level 
tunneling states (TLS) that exist in any crystalline metal. 

The main message of this paper is the following. There exist two 
mechanisms of electron dephasing due to dynamic defects. The first one is 
due to direct inelastic transitions between the levels of the TLS 
leading to the possibility of determining the actual path of the 
electron, and consequently to loss of interference. 
The second one is due to relaxation dynamics of dynamic 
defects which fluctuate due to interaction with the thermal bath. Time 
dependence of the electron scattering crossection due to the defects' 
fluctuations lead to violation of the time-reversal symmetry and, as a 
consequence, to decoherence. To our knowledge, the theory relevant to 
the second mechanism has not been developed. However, there exists a 
temperature interval where this relaxation  mechanism is dominating.     
 
The paper is organized as follows.  Below we will give physical
considerations  
to describe dephasing by  dynamic defects  
which will be then confirmed by a diagrammatic approach, see 
Sec.~\ref{theory}.  In this section the model for electron-TLS 
interaction will be formulated, Sec.~\ref{model}; this model will used 
to calculated the dephasing rate due to \emph{identical} TLSs, 
Sec.~\ref{Cooperon}, and, finally, an average procedure over different 
TLSs will be  
considered, Sec.~\ref{average}. Estimates and discussion will be 
given in Sec.~\ref{estimates}, while the conclusions will be given in 
Sec.~\ref{conclusions}. 
 
\subsection{Qualitative considerations} \label{phys_cons} 

Let us start with a toy model which illustrates the essence of the
physics involved. Then in Sec.~\ref{theory} the results will be confirmed by calculation.

\begin{figure}[h] 
\centerline{
\includegraphics[width=4cm]{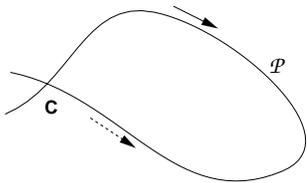}
}
\vspace*{2mm} 
\caption{A closed-loop trajectory. \label{f_1}} 
\end{figure}

Consider the electron motion in a slowly varying potential field $U({\bf 
r},t)$. Let us calculate the phase difference $\Delta\varphi$ 
between the electron waves moving from the same point $C$
along the 
same closed path $\cal P$ clockwise and counterclockwise, see 
Fig.~\ref{f_1}. 
We begin with evaluation of the variation $\Delta S$ of the 
electron's action $S$ due to the time variation of potential $U$. 
We assume that an electron during its motion along the trajectory $\cal 
P$ experiences many 
scattering events against both static 
and dynamic defects, so that the trajectory can be 
approximated by a smooth curve. We have  
\begin{equation} \label{p1} 
S= \int{\bf p}d{\bf r}= \int{\bf p}{\bf n}\, ds 
\end{equation} 
where ${\bf n}={\bf p}/p$ is a unit vector parallel to the tangent to 
the curve $\cal P$, $ds$ is the length element of the curve. This can 
also be written  
$$ 
S=\int ds\sqrt{2m({\cal E}-U)} \, , 
$$ 
${\cal E}=p^2/2m$ being the electron kinetic energy while $m$ is the 
electron effective mass.  Expanding this equation in powers of the 
potential energy $U$ assumed small, one gets 
$$ 
\Delta S=-\int\frac{ds}{v}U(s,t)=-\int dt \, U(s_t,t). 
$$ 
Here $s_t$ is the electron's coordinate on the trajectory parameterized 
by time $t$. So $U$ depends on time both via the space coordinate $s_t$ 
and explicitly.

Let now $t_0$ be the total time of the motion of an electron along 
the loop $\cal P$.
Accordingly, the 
phase variation in the course of a clockwise motion is 
\begin{equation} \label{df-} 
(\Delta\varphi)_+=-\frac{1}{\hbar}\int_0^{t_0} dt\, U(s_t,t) \, , 
\end{equation} 
while for the counterclockwise motion one has 
\begin{equation} \label{df+} 
(\Delta\varphi)_-=-\frac{1}{\hbar}\int_0^{t_0} dt \,U(s_{t_0-t},t) \, . 
\end{equation} 
The dephasing means a non-vanishing phase difference $\Delta\varphi 
\equiv (\Delta\varphi)_+-(\Delta\varphi)_-$. Thus, 
$$ \overline{(\Delta\varphi)^2}=\sum_{\pm}\left[\overline{
    (\Delta\varphi)_\pm^2} -  
\overline{ (\Delta\varphi)_\pm  (\Delta\varphi)_\mp}\right]\, . 
$$ 
Using Eqs.~(\ref{df-}) and (\ref{df+}) one can express the above 
expression through $\int_0^{t_0} \! dt 
\int_0^{t_0}\! dt'\,  \overline {U(s_{t_i},t)U(s_{t'_k},t')} 
$ where $ i,k=\pm, \ 
t_+ \equiv t, \ t_-\equiv t_0-t$. 
We assume that there is \emph{no spatial correlation} between the 
scattering centers, 
$\overline{U(s_t,t)U(s_{t'},t')}\propto\delta(s_t-s_{t'})$, 
that implies 
\begin{eqnarray*} 
\overline {U(s_{t_\pm},t)U(s_{t'_\pm},t')}&\propto& 
\overline {U^2(s,t)}\, \delta (t-t')\, , \\ 
\overline {U(s_{t_\pm},t)U(s_{t'_\mp},t')}&\propto&\overline
{U(s,t)U(s,t_0-t)}\, \delta (t+t'-t_0)\, .  
\end{eqnarray*} 
Using these expressions and introducing the time correlation function of
the time-dependent random potential as  
$$ 
\overline{U(s,t)U(s,t')} \equiv \overline{U^2}f(t-t')\, , \ 
\overline{U^2} \equiv \overline {U^2(s,t)}\, , \ 
f(0)=1\, , 
$$ 
one obtains 
$$ 
\overline{(\Delta\varphi)^2} \propto \overline{U^2} \int_0^{t_0} dt 
\left[1-f(2t-t_0)\right]\, . 
$$ 
If there are several mechanisms responsible for dephasing 
characterized by different coupling strengths and different 
correlation functions the resulting phase variance can be expressed as 
\begin{equation} \label{dr01} 
\overline{(\Delta\varphi)^2} \propto \sum_s  \int_0^{t_0} 
\frac{dt}{\tau_s}\, \left[1-f_s(2t-t_0)\right]\, . 
\end{equation} 
Here we have absorbed the random scattering potential into the partial 
relaxation rates $\tau_{s}^{-1}$. They, as well as 
the correlation functions, depend on the properties of dynamic 
defects. We wish to emphasize that Eq. (\ref{dr01}) demonstrates
the following point indicated above. If the defect has not
relaxed during the time $2t-t_0$ between two acts of scattering
then in spite of non-invariance of the Hamiltonian respective
to the time reversal there is no phase relaxation.


We distinguish two mechanisms of dephasing. The first is connected 
to interactions which cause real transitions between different states of the 
environment. This can be illustrated by the famous double slit experiment. 
If we send electrons at the double slit, it will pass through both 
slits and interfere with itself, creating an interference pattern 
on the screen. Putting detectors to determine which slit the 
electron really passed through will destroy the interference pattern. 
If the interaction with the environment in any way
allows us to  determine the path of the electron, interference is lost. 
The second mechanism of dephasing is related to a change in the state
of the environment due to its own internal dynamics. 
A dynamic environment leads to a difference in a scattering potential
``felt'' by an electron state during clockwise and counterclockwise
motion. As a result, time-reversal symmetry is broken and the
interference pattern decays. 

At this point we would like to compare our description to the one given 
in Ref.~\onlinecite{Stern} where it is proved that the dephasing can 
be described in two equivalent ways. Either you consider the 
change in the electron phase of you consider the change of state of the 
environment, where complete dephasing corresponds to the environment 
being in orthogonal states. The last point of view would imply the 
existence of only the first mechanism of dephasing that we consider. 
We want to emphasize that our second mechanism is not in conflict with 
this, but is a result of our \emph{description} of the process. 
In Ref.~\onlinecite{Stern} the environment is considered as a mechanical 
system evolving according to its own Hamiltonian, whereas we consider the 
environment to be a statistical system at some temperature. That is, we 
calculate the action of the environment on the electrons, but do not 
consider the action of the electrons on the environment. In principle, 
if one were to follow all the complex dynamics of the environment one would 
find that it does indeed evolve into orthogonal states as the electron 
dephases according to the third mechanism, and it would be seen that 
this is only the first mechanism in disguise. However, as the environment 
consists of a macroscopic number of degrees of freedom is it more natural 
to treat it statistically as a thermal bath. In other words, the phase
of an electron state forming a Cooperon trajectory decays due to
\emph{real transitions} in the thermal bath formed by other electrons
assisted by \emph{virtual} processes involving dynamic
defects. In a perturbative approach these processes occur in the
fourth order in the electron-defect coupling constant. In particular,
they do not enter the second-order calculation of defect-enhanced
electron-electron interaction.~\cite{Schwab} However, it will be shown
that they play an important role in dephasing.

A different classification can be made by discriminating between  
the two different  regimes of phase dynamics -- phase \emph{jumps} and
phase \emph{wandering} (or phase diffusion). 
To understand this, let us consider the first mechanism. Consider the case  
of interest for weak localization, that of one electron traveling 
around a closed loop both in the clockwise and counterclockwise direction, 
and interferes with itself after completing a full circuit. 
If we can determine which direction the electron went, we will not 
get interference. If we are unable to do so, it will appear. 
As a detector we use a two level system that is placed at a point on the 
right hand side of the loop, the distance from the starting (and 
ending) point being a fraction $\alpha<\frac{1}{2}$ of the total circumference.
\begin{figure}[h] 
\centerline{
\includegraphics[width=5cm]{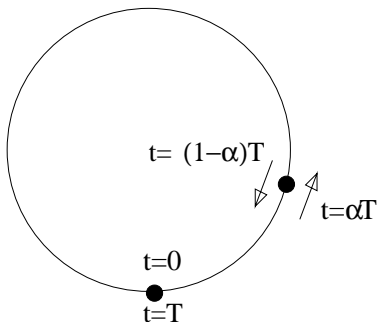}
} 
\caption{Closed loop with TLS detector. \label{fig2}} 
\end{figure} 
The energy splitting of the two level system is $E$, and it starts out in the 
lower state. When the electron passes, it excites the two level system.
We determine the direction of the electron by measuring the time at which 
this happens. The accuracy with which we can make this measurement 
is limited by the uncertainty principle $\Delta t \Delta E > \hbar $.
If $E$ is large there is no problem, and the interference is destroyed 
by a single detection event. This case we call a  phase jump. In the 
opposite case where $E$ is small we can not determine with certainty which 
direction the electron went, and the interference pattern will be 
smeared out, but not lost entirely. In this case we need the interaction 
with a number of two level systems along the path, and the combined 
result of all the detection times can be put together to determine 
the direction. In this case we speak about phase wandering or phase 
diffusion, since the random contributions of the difference two level 
systems makes the electron phase change in a diffusive way.

Let us estimate the dephasing rates in the different cases. 
Consider first the first mechanism where we have 
inelastic electron scattering due to \emph{direct}
transitions between the two TLS's states. We emphasize that the
inelasticity is not essential to this channel of dephasing. The
important point is that there is a real transition between orthogonal
states of the TLS. Since we are not considering degenerate states this
will mean inelastic scattering in our case.  If the energy transfer
$E$ is large enough, the phase relaxation time $\tau_{\varphi}$ is
equal to the typical inelastic relaxation time $\tau_1$, which is a
function of the defect parameters.  The criterion of
``large'' $E$ in this case is given as $E\tau_1/\hbar >> 1$.  For
smaller $E$ one deals with a phase diffusion or \emph{wandering}. To
estimate the dephasing time for this case let us recall that the phase
coherence for any two-level system is conserved during the time $t <
\hbar/E$. While traversing the trajectory during the time $t$ an
electron appears to be coupled with $\bar{N} \sim t/\tau_1$ dynamic
defects. The evolution of the electron wave function due to coupling
with any of these defects is described by a phase factor $\exp ( \pm
iEt/\hbar)$ where $\pm$ corresponds to the sign of the energy
transfer. If $T \gtrsim E$ the probabilities of the both defect states
are almost equal, and the correlation function of the time-dependent
random potential is $f(t)=\cos (Et/\hbar)$, see the calculation later.
The resulting electron
phase shift turns out to be $\bar{N}^{1/2}Et/\hbar$.  Consequently,
the phase relaxation time can be estimated as $\tau_{\varphi} \sim
\hbar^{2/3} \tau_{1}^{1/3}/E^{2/3}$. A similar expression for the
dephasing time has been introduced in Refs.~\onlinecite{aalk} in
connection with decoherence due to quasielastic electron-electron
scattering and in Refs.~\onlinecite{agg,agg1} in connection with
decoherence by low-frequency phonons. In the following we will call
this regime the \emph{phase wandering}.  Summarizing, we can express
the contribution of inelastic processes as
\begin{equation} \label{eq:res001} 
\tau_\varphi^{(1)} =\max\left\{\tau_1, \tau_1^{1/3}(\hbar/E)^{2/3} 
\right\}\, .  
\end{equation} 
 

Moving to the second mechanism, we find that
the simplest way to evaluate this contribution is note that 
$\tau_s$ has a sense of the time at which 
 $\Delta \varphi (t) \approx 1$ provided 
\emph{all} the involved defects  would suffer a transition. It is 
clear that if the phase shift $\delta \varphi$ due to  transition of 
a single defect  is $\gtrsim 1$ then 
 a single TLS is enough to produce the 
dephasing. For $\delta \varphi << 1$ the significant phase 
evolution is possible only with the help of many defects. The actual 
dephasing time, $\tau_{\varphi}$, is also sensitive to 
the defect transition rate $\gamma$. Indeed, the correlation 
function $f(t)$ for statistically independent defects is expected to have 
a form $f(t) = e^{- 2\gamma |t|}$ (this form will be supported by the 
calculations in Sec.~\ref{Cooperon}). If $\gamma \tau_3  \gtrsim 1$, 
then with a help of  
Eq.~\ref{dr01} one obtains 
$\overline{ (\Delta \varphi)^2} = t_0/\tau_3$ (for the reasons which 
 will be clear later we ascribe the subscript 3 for the relaxation 
 mechanism) .  If $\gamma \tau_3  << 1$, 
 one has  $ \overline{ (\Delta \varphi)^2} \sim \gamma t_0^2/\tau_3 
 $. We observe that there is a phase wandering regime also for this 
 relaxation mechanism. Again,  
defining $\tau_{\varphi}$ as a time at which $\overline{ (\Delta 
\varphi)^2} \sim 1$ one has 
\begin{equation} \label{eq:rel001} 
\tau_\varphi^{(3)}=\max\left\{\tau_3, (\tau_3/\gamma)^{1/2}\right\}\, . 
\end{equation}

One notes that in course of the above considerations we exploit the 
additions to the electron phase acquired by an electron in course 
of traversing of the potential induced by the TLSs.  For each  TLS the 
corresponding contribution can be estimated as $\tilde{ 
U}\delta r /v_F$, where  $\tilde{U}$ is the potential magnitude, $\delta 
r$ is the potential spatial scale, while $v_F$ is the Fermi 
velocity. An important note should be made in this concern. Since 
the trajectory in Eq.~(\ref{p1}) and in the following ones is 
considered to be given and the positions of the scatterers are 
expected to be along the electron trajectory, the phase addition 
mentioned above is, strictly speaking, beyond the Born 
approximation for the electron scattering. Indeed, in the Born 
approximation the scattering amplitude is real at least for 
symmetric scattering potentials. 
However it is possible to describe the phase relaxation even 
within the framework of the Born approximation if one 
takes into account that the ``centers of gravity'' of the two TLS 
states are spatially separated by some vector 
$\ba$ (which is an inherent feature of the model 
suggested in Ref.~\onlinecite{kr}). In this case the phase 
variation due to a transition within the $i$-th defect is simply 
given as $\delta \varphi \sim (\bp \cdot  \ba)/\hbar $. 
Correspondingly, the estimate for the proper rates in 
Eq.~(\ref{dr01}) is $\tau_{1,3}^{-1} \approx \tau_{e,\text{d}}^{-1}(p_F 
a /\hbar)^2 $ where $\tau_{e,\text{d}}^{-1}$ is a typical 
\emph{elastic} relaxation rate due 
to dynamic defects.~\cite{Black,Imry} 
  
The previous estimates are relevant to a set of defects having 
\emph{identical} parameters. However, in real systems the defect 
parameters are scattered, and one has to perform a proper average. As 
we will demonstrate, for a realistic model the phase wandering regime 
turns out to be important. To our knowledge, this fact has not been 
appreciated in the previous papers dealing with defect-induced 
decoherence. 
   
In the following sections we will give a more formal derivation of 
the dephasing rate using the Green function method. It will permit us to 
consider not only the limiting cases but any relations 
between various times of relaxation. In the Appendix~\ref{rtn} we will
map the results for the relaxation-dynamics contribution to a simple
model of short-range defects hopping between two states separated some
distance in real space. This model is often used to interpret results
on the so-called random telegraph noise observed in nanostructures.
 
\section{Theory} \label{theory} 
 
\subsection{The model} \label{model} 
 
The dephasing  mechanism is based on the assumption that in 
any crystalline metal there exist dynamic defects of a special 
type. These defects are tunneling states which are described by the 
Hamiltonian 
\begin{equation} \label{tls01} 
{\cal H}_{\text{d}}=(\Delta \,  \sigma_3 - \Lambda \, \sigma_1)/2 
\end{equation} 
where $\Delta$ is the diagonal level splitting, $\Lambda$ is the 
tunneling amplitude, while $\sigma_i$ are the Pauli matrices. The 
tunneling amplitude $\Lambda$ describes the tunneling between the 
interstitial positions while the spread of $\Delta$ is determined 
by (mesoscopic) disorder around the mobile defect. Consequently, 
we will assume that the distribution of $\Lambda$ is narrow and it 
is centered around some value $\Lambda_0$. As we will see, one can 
expect smooth temperature dependence of dephasing at $T \gtrsim 
\Lambda_0$. The above model has been proposed and successfully 
exploited in Ref.~\onlinecite{kr} to interpret zero-bias anomalies 
observed in metallic point contacts. Note that it  differs from 
the well-known TLS model in amorphous metals~\cite{Black} where 
the distribution of $\log \Lambda$ is assumed to be uniform, 
however resembles the TLS model for crystalline materials 
suggested by Phillips \cite{Phillips1} to describe acoustic 
experiments in crystalline Si. 
 
Consider now spinless electrons which scatter 
against tunneling defects with the Hamiltonian~(\ref{tls01}). 
The total 
Hamiltonian then can be expressed in the form 
\begin{equation} \label{tH} 
\tilde{\cal H}={\cal H}_{\text{d}}+\sum_{\bp} \epsilon_{\bp} c^+_{\bp}
c_{\bp}  
+{\cal H}_{\text{int}} 
\end{equation} 
where 
\begin{equation} \label{iH} 
{\cal H}_{\text{int}}= 
\frac{1}{2} \sum_{\bp \bp' n} \left({\hat  {\mathit 1}}{\tilde
    V}^{+}_{\bp \bp'}+\sigma_3  
{\tilde V}^{-}_{\bp \bp'} \right) 
c^+_{\bp} c_{\bp'}\, e^{i(\bp -\bp')\cdot \br_n/\hbar} \, . 
\end{equation} 
Here $\hat {\mathit 1}$ is the unit matrix, $V^{\pm}$ represent the components 
of a short-range defect potential, while $\br_n$ is the coordinate of the 
$n$-th defect. The Hamiltonian (\ref{iH}) is equivalent to the 
assumption that the electron scattering amplitudes are ${\tilde 
  V}^{+}\pm {\tilde V}^{-}$ in the ``left'' and the ``right'' 
defect positions, respectively. Estimates for  ${\tilde V}^{+}$ and  ${\tilde 
  V}^{-}$ are given in Refs.~\onlinecite{Black,Imry}. 
After the transform which makes 
$\tilde{\cal H}_{\text{d}}$ diagonal we arrive at the Hamiltonian 
\begin{eqnarray} \label{dh} 
{\cal H}&=& \frac{1}{2}\sum_n E_n \, \sigma_3+\sum_{\bp} 
\epsilon_{\bp} c^+_{\bp} c_{\bp} 
+\frac{1}{2} \sum_{\bp \bp'n} \left\{{\hat {\mathit 1}}V^{+}_{\bp 
    \bp'} \right. 
\nonumber \\ 
&&\left.+ \left(\frac{\Lambda_n}{E_n} \sigma_1 +\frac{\Delta_n}{E_n} 
    \sigma_3\right)V^{-}_{\bp \bp'}\right\} 
c^+_{\bp} c_{\bp'} \, e^{i(\bp -\bp')\cdot \br_n/\hbar}\, , 
\end{eqnarray}
where $E_n=\sqrt{\Delta_n^2+\Lambda_n^2}$.
One observes that there are two processes of electron-defect 
interaction described by the items proportional to $\sigma_1$ and 
$\sigma_3$, respectively. They correspond to the two mechanisms 
discussed above and described by Eqs.~(\ref{eq:res001}) and (\ref{eq:rel001}).
Now we proceed to more formal calculations in which the relaxation 
time $\tau_1$ and $\tau_3$ will be specified.

\subsection{Quantum contribution to conductance}\label{Cooperon} 
 
\begin{figure}[h] 
\centerline{
\includegraphics[width=7cm]{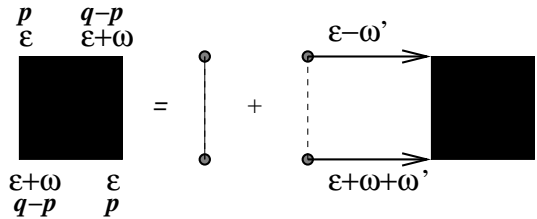}
} 
\caption{The equation for the Cooperon. \label{f-c1}} 
\end{figure}

The object which we will consider is the weak localization correction 
to the conductivity, $\delta \sigma$, which for the case of a 
short-range scattering potential can be expressed through 
the electron Green's functions $G_{R/A}$ in the form\cite{agg} 
\begin{eqnarray} \label{con1} 
&&\delta \sigma = \frac{e^2}{m^2d}\int (dp)\,(dq)  p^2 \int \left(-\frac{dn}{d 
\varepsilon}\right)\, \frac{d \varepsilon}{2\pi} \int \frac{d \omega}{2\pi} 
\nonumber \\ &&\  \times 
 G_R(\varepsilon,{\bp})G_A(\varepsilon,{\bp})\, F(\varepsilon, 
\omega, {\bp},{\bq - \bp}) 
\nonumber \\ 
 &&\quad \quad \times G_R(\varepsilon+ 
\omega,{\bq}-{\bp})G_A(\varepsilon + \omega,{\bq}-{\bp)} \, . 
\end{eqnarray} 
 Here $d$ is the dimension of the problem, $(dp)\equiv d^dp/(2\pi 
\hbar)^d$, $n(\varepsilon)$ is the Fermi function, while $F(\varepsilon, 
\omega, {\bp},{\bp_1})$ is a two-particle Green's function specific for 
the problem under consideration. The above expression describes the 
main contribution to the conductivity which arises from the region 
$\omega \tau, \, q \ell \ll 1$ where $\tau$ and $\ell$ are the total 
relaxation time and length, respectively. The function $F(\varepsilon, 
\omega, {\bq}, {\bp})$ can be represented as a sum of the 
maximally-crossed diagrams (the so-called Cooperon) which is a sum of 
a ladder in the particle-particle channel. The Cooperon satisfies the 
Dyson equation shown in Fig.~\ref{f-c1}. 
In this figure, the Cooperon is drawn as a filled square, thick lines 
with arrows correspond to the Green's functions averaged over the 
defect position, as well as over the states of the thermal bath, while 
dotted lines represent propagators for electron scattering against 
dynamic defects. Since the interaction Hamiltonian (\ref{dh}) contains 
items of three types, the propagator consists of a sum of three 
terms. Each propagator can be expressed as a loop graph where dotted 
lines represent Green's functions for a dynamic defect. 
\begin{figure}[h] 
\centerline{
\includegraphics[width=3cm]{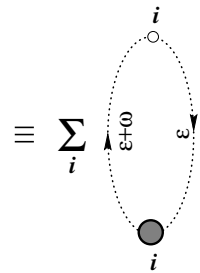}
} 
\caption{Schematic representation of the defect propagator. \label{f-c2}} 
\end{figure} 
To express the propagators in an analytical form we will employ 
the technique developed by Abrikosov.\cite{abric} According to 
this technique, a two-level system describing the dynamic defect 
is interpreted as a pseudo-Fermion particle with the Green's 
function 
\begin{equation} 
g_\pm(\epsilon)=(\epsilon \mp E/2- \lambda + i \delta)^{-1}\, , 
\end{equation} 
where $\lambda$ is an auxiliary ``chemical potential'' which 
afterwords will be tended to infinity. This trick allows one to remove 
extra unphysical states which appear since the Fermi operators have 
more extended phase space that the spin ones.\cite{abric,maleev} As a 
result, the Matsubara technique can be effectively used, and after a 
proper analytical continuation and the limiting transition $\lambda 
\rightarrow \infty$ the quantity $\lambda$ drops out of all the 
expressions. As a result, the retarded propagator describing the inter-level 
transitions in the defect can be expressed as 
\begin{equation} \label{D1} 
{\cal D}_1^R (\omega)= -\tanh \frac{E}{2T}\left( \frac{1}{\omega -E + i 
\delta}- \frac{1}{\omega +E + i 
\delta}  \right)\, . 
\end{equation} 
Here $\delta$ is the adiabatic parameter, $\delta \rightarrow +0$. 
The propagator describing electron-assisted transitions has the form 
\begin{equation} 
{\cal D}_3^R (\omega)=\frac{1}{T \cosh^2 (E/2T)} \, \frac{2i \gamma}{\omega 
+ 2i \gamma }\, . 
\label{eaa} 
\end{equation} 
Here 
\begin{equation} 
\gamma (\Lambda, E)= \left(\frac{\Lambda }{E}\right)^2\! \gamma_0(E)\, , \quad 
\gamma_0(E)=\chi E \coth \frac{E}{2T} \, , 
\label{gaa} 
\end{equation} 
where $\chi =0.01-0.3$ is dimensionless constant dependent on the 
matrix element $V^{(1)}$ where $\gamma_0 (E)$ has 
the meaning of \emph{maximum} hopping rate for the systems with a 
given interlevel spacing.~\cite{Black}

For the elastic component $\propto \hat {\mathit 1}$ we shall use a trick 
which will allow us to consider the elastic channel in a unified way 
with the inelastic ones. Namely, to keep proper analytical properties 
of the retarded Green's function we define the elastic propagator as 
\begin{equation} 
{\cal D}_0^R (\omega)=\frac{\nu}{2T} 
\left(\frac{1}{\omega +\nu + i \delta } - \frac{1}{\omega -\nu + i 
\delta } \right) \, . 
\label{faa} 
\end{equation} 
At the final stage, the limiting case $\delta 
\rightarrow 0, \ \nu \rightarrow 0$ should be calculated. Note that 
the factor $T^{-1}$ will be canceled by the Planck function $N_0(\omega)$ 
which will appear in course of derivation of the equation shown in 
Fig.~\ref{f-c1}.  The 
physical reason of this cancellation is that the elastic impurity 
scattering is temperature-independent. 
Note that the propagators do not include the electron-defect coupling 
constant, hence each propagator should be multiplied by 
$|W^{(i)}|^2$ where $W^{(0)}=V^+$, $W^{(1)}=(\Lambda/2E)V^-$, 
$W^{(3)}=(\Delta/2E)V^-$ . If there are
additional static short-range defects their contribution modifies 0th 
propagator by the replacement $|W^{(0)}|^2 \rightarrow  
|W_s|^2+|W^{(0)}|^2$ where $W_s$ is the contribution of static defects.
 
The equation shown in Fig.~\ref{f-c1} has been analyzed following the 
procedure of analytical continuation of Matsubara Green's 
function\cite{agg} with making use of analytical properties of 
two-particle Green's functions.\cite{maleev1} The resulting equation 
for $F(\varepsilon,\omega,\bp, \bq-\bp_1)$ has the form 
\begin{eqnarray} 
&&F(\varepsilon, \omega,\bp, \bq -\bp)= {\cal D} ( \omega ) \nonumber \\ 
&&-\int \frac{(dp')\, d \omega' }{2\pi i}F(\varepsilon, \omega',\bp', 
\bq - \bp 
){\cal D}(\omega - \omega ') \nonumber \\ 
&&\quad \times G^R(\varepsilon + \omega -\omega', {\bf p}' ) \, 
G^A(\varepsilon + 
\omega', \bq -\bp')\nonumber \\&& \quad \quad \times 
\left[N_0(\omega') -N_0(\omega' - 
\omega) \right] \, . 
\label{eco} 
\end{eqnarray} 
Here $(dp) \equiv  2\, d^2p/(2 \pi \hbar)^2 \equiv \rho\, d 
\varepsilon_\bp \, d \theta/2\pi$, $\theta$ is the angle with the $x$ axis
and we write all formulas for the most interesting case of a two dimensional 
system. 
\begin{equation} 
\begin{split}
{\cal D} ( \omega ) 
 \equiv &|W_s|^2n_s\left[{\cal D}_0^R( \omega )- {\cal D}_0^A(\omega )\right]\\
 &+ \sum_{i} |W^{(i)}|^2n_d\left[{\cal D}_i^R( \omega )- {\cal D}_i^A(\omega )
   \right]\, . \label{d1} 
\end{split}
\end{equation} 
Equation~(\ref{eco}) describes the dominant contribution provided 
the sum of the incoming momenta, $q$, is small: 
\begin{equation} 
q \ell \ll 1 \, . 
\label{iaa} 
\end{equation} 
Here $\ell=v_F\tau$ is the electron  mean free path, while $\tau$ is the 
electron life time,  
\begin{equation} 
\tau^{-1}= \tau_e^{-1}+\tau_1^{-1}+ \tau_3^{-1} \, . 
\label{kaa} \end{equation} 
Here we introduce the elastic relaxation rate as a sum of the 
contributions of static and dynamic defects, 
$\tau_e^{-1}=\tau_{e,s}^{-1} +\tau_{e,d}^{-1}$ with  
$$\tau_{e,s}^{-1} 
=2\pi \rho n_s | V_s|^2/\hbar \, , \quad \tau_{e,d}^{-1} 
=2\pi \rho n_d|V_d^+|^2)/\hbar\, ,$$  
and a typical inelastic 
relaxation rate as 
\begin{equation} \label{eq:inr} 
\tau_i^{-1}=2\pi \rho n_d  |V_d^-|^2/\hbar\, , \quad \tau_{e,d}/\tau_i 
\approx (p_F a /\hbar)^2 \lesssim  1 \,. 
\end{equation} 
Here $n_s$ is the 
concentration of static defects while $n_d$ is the concentration of 
dynamic ones,  
$\rho$ is the electron density of states. In principle we now 
have two sets of relaxation rates. From the interaction vertices we get 
$$ 
\tau_{1}^{-1} =\tau_i^{-1}(\Lambda/2E)^2\, , \quad 
\tau_{3}^{-1}=\tau_i^{-1}(\Delta/2E)^2\, , 
$$ 
while the rates appearing in (\ref{kaa}) arises in the evaluation of the 
self energy diagrams as shown in appendix A, and are given by 
$$ 
\tau_{1}^{-1} =\tau_i^{-1}(\Lambda/2E)^2{\cal G}_1(\varepsilon)\, , \quad 
\tau_{3}^{-1}=\tau_i^{-1}(\Delta/2E)^2{\cal G}_3(\varepsilon)\, . 
$$ 
 The functions ${\cal G}_{1,3}$ are 
discussed in appendix A. We show that they decay exponentially at 
$E \gg T$, and only the regions with $\varepsilon \le T$ are 
important. Since we only are interested in this region, we will neglect the 
energy dependence of these functions, and put them to 1 in the following. 
The two sets of relaxation rates will then be the same. 
 
To analyze Eq.~(\ref{eco}) it appears convenient to transform it 
to the form similar to the Boltzmann equation for an electron 
diffusion. For this let us take into account that at small $q$ and 
$\omega$ the product of the Green functions in the integrand is a 
sharp function centered at $\varepsilon = 
\varepsilon_{\bp'}=\epsilon_F$ where $\epsilon_F$ is the Fermi 
level. Thus it is natural to assume that $F(\varepsilon, \omega, 
\bp', \bq - \bp_1 )$ depends only on $q$, $\omega$ and the product 
$\bq \cdot \bv'$. Having that in mind we first integrate over 
$\varepsilon_{\bp'}$ and make use of the inequalities 
$p_F\ell/\hbar \gg 1, \ \hbar \omega \ll T$ which we assume to be 
met.  

 The result can be expressed in terms of a new function 
\begin{equation} 
{\cal F}(\varepsilon,\bq, \omega ) \equiv \frac{F(\varepsilon, \omega,\bp, 
\bq - \bp 
)}{\omega (1- i\tau 
{\bf qv} )} 
\label{jaa} 
\end{equation} 
where ${\bf v}$ is the electron velocity. Here we assume that 
$\varepsilon \le T$ and omit the variable $\varepsilon$. 
 
Following the procedure 
described in Ref.~\onlinecite{agg} we express the equation 
for $\cal F$ in the form: 
\begin{eqnarray} 
&&(1+ Dq^2 \tau ){\cal F}(\varepsilon,\bq, \omega )=
 \frac{{\cal D} (\omega)}{4\pi\rho \omega} \nonumber \\ 
&& - T\int 
\frac{d \omega'}{(2\pi i)(\omega -2 \omega'+i/2\tau) }{\cal 
F}(\varepsilon, \bq, \omega' )\,  \frac{{\cal D}
 (\omega-\omega')}{\omega-\omega'}\,.  
\label{maa} 
\end{eqnarray} 
Here $D=v_F\ell/d$ is the diffusion constant.
Transforming Eq.~(\ref{maa}) to the time representation with respect 
to $\omega$ we obtain 
\begin{eqnarray} 
(1&+& Dq^2 \tau){\cal F}(\varepsilon,\bq,t )= \frac{\Phi(\varepsilon,t)}{2\tau T\rho}\nonumber \\ 
&&+\int_{-\infty}^t 
\frac{dt'}{\tau}e^{(t'-t)/\tau}{\cal F}(\varepsilon,\bq,t')\Phi(\varepsilon,2t-t')\, , 
\label{teq} \end{eqnarray}
Here we denote
\begin{equation}
\Phi (\varepsilon,t) \equiv \frac{\tau}{\tau_e} + \frac{\tau}{\tau_1} \cos 
\frac{Et}{\hbar} + \frac{\tau}{\tau_3} e^{-2\gamma t/\hbar}  \, . 
\label{oaa} 
\end{equation} 
Here the limiting transition $\nu \rightarrow 0$ has been already
done.  Note that the function $\Phi(\varepsilon,t)$ depends on the
energy variable $\varepsilon$ through the relaxation times
$\tau_e$, $\tau_1$ and $\tau_3$. In the following we omit the variable $\varepsilon$ in all
the functions keeping in mind that the relaxation rates are
energy-dependent, see Appendix A. Also in writing the expression for 
$\Phi(\varepsilon,t)$ we have assumed that $E\ll T$. For $E>T$ it decays 
to 1 which means that defects with $E>T$ do not contribute to 
the dephasing, see the discussion below Eq.~(\ref{4ab})
 
Equation~(\ref{teq}) can be solved exactly. The solution is based on the 
relation between the kernel of the integral equation, 
\begin{eqnarray*} 
&&{\cal K} (t, t')=(Dq^2\tau -1)e^{(t'-t)/\tau} \left[1 - 
\lambda(2t-t') \right] ;\nonumber \\ 
&&\lambda(t)\equiv \frac{\tau}{\tau_1}\left(1 - 
\cos\frac{Et}{\hbar} \right) 
+\frac{\tau}{\tau_3}\left(1 - e^{-2\gamma t/\hbar}\right) 
\end{eqnarray*} 
and its resolvent, $\cal R$, defined by the integral equation 
\begin{equation} 
\int_{t_1}^t \frac{dt'}{\tau} {\cal K} (t_1, t') {\cal R}(t',t)= {\cal 
K}(t_1,t)+{\cal R}(t_1,t) \, . 
\label{saa} 
\end{equation} 
The relationship has the form\cite{evans} 
\begin{equation} 
{\cal F}(t,\bq)=\Phi(t) +\int_{-\infty}^{t} \frac{dt'}{\tau} {\cal 
R}(t,t') \Phi(t')\, . 
\label{raa} 
\end{equation} 
If one can construct a differential operator of the form 
$ 
{\hat {\cal L}}_{t_1} \equiv \sum_j a_j(t_1) (d^j/dt_1^j) 
$ 
such that 
\begin{equation} 
{\hat {\cal L}}_{t_1} {\cal K}(t_1,t)=0 
\label{uaa} 
\end{equation} 
for any $t$, then the integral equation (\ref{raa}) is reduced to the 
differential equation (\ref{uaa}) for a fixed $t$. That can be 
directly checked applying the operator ${\hat {\cal L}}_{t_1}$ to 
relation (\ref{saa}). The boundary 
conditions corresponding for $t_1 \rightarrow t$ can be extracted from 
the integral relation (\ref{saa}) and its derivatives with respect to 
$t_1$ at  $t_1 \rightarrow t$. 
 
The results have the simplest form at $\tau_e \ll \tau_1,\tau_3, 
\hbar/\gamma$. Then one choose 
$$ 
{\hat {\cal L}}_{t_1} = \left(\tau\frac{d}{dt_1}+1 \right)^3\, . 
\label{vaa} 
$$ 
{}From (\ref{saa}), the differential equation for the resolvent ${\cal 
R} (t_1,t)$ acquires the form 
\begin{eqnarray} 
\left[\tau^3 (d^3/dt^3)\right. & + &  \tau^2 (2+\lambda_1)(d^2/dt^2)  
\nonumber \\   & +&  \left. 
\tau (1+ 2 \lambda_1) (d/dt) + \lambda_1 \right]{\cal R}(t,\theta)=0 \, . 
\label{waa} 
\end{eqnarray} 
Here $\lambda_1=\lambda+Dq^2\tau$. 
Since the phase relaxation is a slow process with respect to the scale 
$\tau$ the equation (\ref{waa}) has small coefficients at senior derivatives 
which makes useful the WKB approximation. Consequently, the physical solution 
can be sought in the form 
$ 
{\cal R}(t,\theta) \propto \exp[\varphi (t)/\tau] 
$ 
where $\varphi (t)$  satisfies the equation, 
$ 
\left(
{\dot \varphi} +1 \right)^2\left[ 
{\dot \varphi} 
+\lambda_1 (t) \right]=0 
$.  
Since $\lambda (t) \ll 1$, the WKB solution corresponds to the 
equation 
\begin{equation} 
{\dot \varphi}
+\lambda_1 (t) =0\, . 
\label{zaa}  
\end{equation} 
The boundary condition to Eq.~(\ref{zaa}) can be extracted from the relation 
$ {\cal R}(t,t)=0 
$. 
In this way, we obtain the quasiclassical solution in the form 
\begin{equation} 
{\cal R}(t,t_1)= \exp \left(- \int_{t_1}^t\frac{dt'}{\tau}\, \lambda_1 
(t') \right)\, . 
\label{3aa} 
\end{equation} 
Now we substitute Eq.~(\ref{3aa}) in the expression (\ref{raa}) to 
obtain the final expression for the Cooperon $\cal F$. The first item 
in Eq.~(\ref{raa}) is the contribution of lowest order scattering and 
it should be neglected in the diffusion approximation. Here we analyze 
the quantum contribution to the static conductance, so only ${\cal F} 
(0)$ is important. As a result, we obtain 
\begin{eqnarray} 
{\cal F}(0,\bq)&=&\int_{-\infty}^0 \! \frac{dt'}{\tau}\, \Phi(t')\, 
e^{\vartheta(t')} \, , \label{4aa}\\ 
\vartheta (t)&=&
Dq^2t + 
\left[\frac{t}{\tau_1}- \frac{\sin(Et/\hbar)}{E\tau_1 / \hbar  } 
\right] \nonumber \\ 
&&+\left[\frac{t}{\tau_3}- \frac{\hbar }{2\gamma \tau_3}\big( e^{2\gamma t/\hbar}-1\big)  \right], \quad t<0 \, . 
\label{4ab} 
\end{eqnarray} 
An important feature of Eq.~(\ref{4ab}) is that if one neglects the 
processes in which the defect changes its state then the 
\emph{dephasing is absent}. Indeed, putting $\tau_1=\tau_3=\infty$ we 
get $\Phi (t)=1$ and ${\cal F}(\bq) = (Dq^2)^{-1}$.  This results in 
logarithmic divergence of the conductance in the 2D case. Another important 
feature is that at small time $t$, which has a physical meaning of the 
time difference for the collision act for clockwise and 
counter-clockwise partial waves, no linear in $t$ term is originated 
by inelastic processes. Physically it means that no dephasing takes 
place if the scattering defect had no enough time to change its 
state. Therefore, the dephasing appears proportional to the 
probability for the defect to escape the state in which it has been 
registered by one partial wave. 
 
 
Substitution Eq.~(\ref{4aa}) into Eq.~(\ref{jaa}) and then into 
Eq.~(\ref{con1}) we obtain for 2D case the 
expression for the quantum contribution to the conductance. Since only 
$|\varepsilon| \lesssim T \ll \epsilon_F$ are important one can neglect 
$\varepsilon$-dependence of the relaxation rates and put $\varepsilon=0$ 
in the expressions for the these quantities. As a result, one arrives
at the well-known expression,  
$$ 
\delta \sigma =- \frac{e^2}{2 \pi^2 \hbar} \ln \frac{\tau_\varphi}{\tau} 
$$ 
where $\tau_\varphi$ is defined according to the equation 
\begin{eqnarray} 
\ln \frac{\tau_\varphi}{\tau} &\equiv& \int_{1}^\infty \! \frac{d \eta}{\eta} 
e^{-\Gamma_1(\eta,E,\Lambda)-\Gamma_3(\eta,E,\Lambda)}\, ,  
 \label{6aa}\\ 
\Gamma_1(\eta,E,\Lambda)&=&\frac{\tau}{\tau_1}\left[\eta -
  \frac{\sin(\eta\,  
E\tau/ \hbar ) }{E\tau/ \hbar}  \right] 
\, , \nonumber \\ 
\Gamma_3(\eta,E,\Lambda)&=&\frac{\tau}{\tau_3}\left[\eta 
- \frac{\hbar }{2\gamma \tau}\big(1-e^{-2\eta \gamma \tau/ \hbar 
}\big) \right] \, , \nonumber 
\end{eqnarray} 
where $\eta=t/\tau$.
This equation is obtained by the integration over $\bq$.

We can now recover our estimates (\ref{eq:res001}) and (\ref{eq:rel001})
by the approximate value of the integral 
\[
 I = \int_1^\infty\frac{d\eta}{\eta}e^{-\alpha\eta^n}
\]
in the case where $\alpha\ll1$. We split the integral at the point 
$\eta^*$:
\[
 I = \int_1^{\eta^*}\frac{d\eta}{\eta}e^{-\alpha\eta^n} 
    + \int_{\eta^*}^\infty\frac{d\eta}{\eta}e^{-\alpha\eta^n}.
\]
Defining $\eta^*$ such that $\alpha{\eta^*}^n=1$, we have $\eta^*\gg1$. 
The integrand will be very small in the last integral, and we 
neglect this. In the first integral we put the exponent equal to 0, 
and get 
\begin{equation}\label{int}
 I \approx\ln\eta^*=\ln\alpha^{-1/n}.
\end{equation}

Expression (\ref{6aa}) depends upon two dimensionless quantities. The 
first one is $E t/\hbar$. As
follows from Eq.~(\ref{int}), the typical value of $t$ is
$\tau_\varphi$.   This parameter determines the efficiency of the 
first mechanism of dephasing, that of direct transitions of the TLS states.
If $E \tau_\varphi/ \hbar \ll 1$ we are in the regime of phase
wandering, and we can  
expand $\Gamma_1$ in Eq.~(\ref{6aa}) in powers of this parameter up to 
the lowest order. However, if $E\tau_\varphi/ \hbar \gtrsim 1$ we have
the case of  
phase jumps, and we can neglect the sine term in $\Gamma_1$. These 
expansions are given in the equations (\ref{eq:gamma101}) and 
(\ref{eq:gamma102}) below. 
In both cases we can use the formula (\ref{int}) to   
arrive at the estimate (\ref{eq:res001}). 

The second dimensionless parameter is $\gamma \tau_3/ \hbar$. It describes 
the effect of the second mechanism of dephasing arising from the $\sigma_3$
vertex. The physical explanation is that the dephasing occurs only if 
the partial waves meet the scatterer in different states. 
Expanding in small [phase wandering, Eq.~(\ref{eq:gamma201})] and large 
 [phase jumps, Eq.~(\ref{eq:gamma202})] values of this parameter we
arrive at the estimates (\ref{eq:rel001}). 
 
In addition there is also the dimensionless parameter $\gamma \tau_1/ \hbar$
which will control the effect of the second mechanism acting through the 
$\sigma_1$ vertex. This has been neglected in the above calculations
since the inequality $\gamma \ll E/\hbar$ is met.

If the estimates (\ref{eq:res001}) and (\ref{eq:rel001}) have 
different orders of  
magnitude then the shortest one is effective. However, 
$$1/\tau_\varphi \ne 1/\tau_\varphi^{(1)} +1/\tau_\varphi^{(3)}$$ 
since  $\Gamma_1$ and  $\Gamma_3$ depend on time in  different ways. The 
most clear manifestation of this fact is seen in the magnetic field dependence 
of the quantum contribution.

\subsection{Average over different dynamic defects} \label{average} 
 
To calculate the quantum contribution to conductance one has to 
sum over different dynamic defects.  In the previous 
considerations we have assumed that all dynamic defect have the 
same interlevel distance, $E$, and the same hopping rate, 
$\gamma$. Consequently, the summation over different defects has 
been allowed for  by the factor $n_d$ in the expressions for the 
relaxation times $\tau_i$.   However, in realistic 
systems both $E$ and $\gamma$ can be distributed over a significant range. 
Since the number of dynamic defects at a typical electron 
trajectory is assumed to be large the summation over different 
defects can be replaced by a proper average. To calculate the 
latter it is necessary to specify the distribution function 
$\cP(E,\gamma)$ which we assume to be normalized to 1. To specify 
this function, let us come back to the effective Hamiltonian 
(\ref{tls01}).  Since  $\Delta$ is determined by the defect's 
neighborhood while $\Lambda$ is determined by the distance between 
two metastable states it is natural to assume $\Delta$ and 
$\Lambda$ to be  \emph{uncorrelated}, 
$\cP(\Delta,\Lambda)=\cP_\Delta (\Delta)  \cP_\Lambda(\Lambda)$. 
 
Below we will discriminate between two model distributions. The 
first one will be referred to as the ``glass-model'' 
(GM).~\cite{ahvp,Phillips} According to this model the distribution 
of $\Delta$ is assumed to be smooth, $\cP_\Delta = \cP_0$. Since 
the tunneling integral $\Lambda$ is an exponential function of the 
distance between the potential minima and the latter is smoothly 
distributed, it is assumed that $\cP_\Lambda \propto 
\Lambda^{-1}$.  Within this model it is natural to choose the 
interlevel splitting $E$ and the quantity $p\equiv (\Lambda/E)^2$ 
as independent parameters. Since $\gamma \propto p$, 
equation (\ref{gaa}) can be rewritten as  $ 
\gamma =p\,\gamma_0 (E)$. Consequently, the GM results in the 
\emph{exponentially-broad} distribution of relaxation rates. 
Furthermore, to keep the distribution normalized, we introduce 
cut-off $p_{\min}(E)=\gamma_{\min}(E)/\gamma_0 (E) \ll 1$ and 
assume $\cL \equiv \ln (1/p_{\min})=\ln(\gamma_0/\gamma_{\min})$ to 
be energy-independent. A cut-off energy in the smooth distribution 
of $E$ at some $E^*$ is also assumed. As a result, we get the 
distribution 
\begin{equation} \label{gm} 
\cP_{GM}(E,p)=\frac{\Theta(E^*-E)}{E^*\cL}\, \frac{1}{p\sqrt{1-p}}\, . 
\end{equation} 
 
Another model which we will call the ``tunneling-states-model'' 
(TM) is more appropriate for crystalline materials. There the 
tunneling integrals $\Lambda$ is determined by the crystalline 
structure and are almost the same for all dynamical defects. On 
the other hand, the parameter $\Delta$ is determined by long-range 
interactions, and it is assumed distributed smoothly within some 
band, cf. with Ref.~\onlinecite{kr}. Then 
\begin{equation} \label{dm} 
\cP_{TM}(E, \Lambda)=\frac{\Theta(E^*-E)}{E^*}\, 
\frac{E}{\sqrt{E^2-\Lambda_0^2}}\, 
\delta(\Lambda - \Lambda_0)\, . 
\end{equation} 
In the following we will assume that the dynamical defects are characterized 
by $\Lambda_0 \ll T$. 
To calculate the total contribution of the dynamical defects in the 
case when their parameters are random one has to replace $\Gamma_i$ in 
Eq.~(\ref{6aa}) by the averages ${\bar \Gamma}_i(\eta)=\int dE\, d\Lambda\, 
\Gamma_i(\eta,E,\Lambda)$. Below we will discuss in detail only the 
tunneling-states-model which seems to be more appropriate for 
crystalline materials. 
 
Let us discuss the contribution of direct transitions and relaxation 
separately. 
 
\paragraph{Contribution of direct (resonant) transitions.}

The item $\Gamma_1$, responsible for the direct transitions, is 
proportional to $|W_1|^2= (\Lambda/E)^2|V^{(1)}|^2$. Since 
$\cP_{TM}(E, \Lambda) \propto \delta\, (\Lambda - \Lambda_0) $ the 
integral over $\Lambda$ yields 
the factor $\Lambda_0^2/E\sqrt{E^2-\Lambda_0^2}$. 
Using (\ref{int}), 
the quantity $\tau_\varphi^{(1)}$ is estimated from the expression 
$\bar{\Gamma}_1\left(\tau_\varphi^{(1)}/\tau\right) =1$. 
The following calculation depends on the relationship between $E$ and 
$\tau_\varphi$. At $E\tau_\varphi/\hbar \ll 1$ 
one can expand the expression for $\Gamma_1$ as 
\begin{equation} 
\label{eq:gamma101} 
\Gamma_1(\eta,E,\Lambda) \sim (\Lambda/\hbar)^2 (\eta \tau)^3/\tau_i \, , 
\end{equation} 
while at $E\tau_\varphi/\hbar \gg 1$ 
\begin{equation} 
\label{eq:gamma102} 
\Gamma_1(\eta,E,\Lambda) \sim (\Lambda/E)^2 \eta \tau/\tau_i \, . 
\end{equation} 
To estimate $\bar{\Gamma}_1 (\eta)$ let us introduce the energy 
splitting $T_\Lambda$ at which $E \tau_\varphi/\hbar=1$. The meaning of this 
is that a TLS with an energy splitting less than $T_\Lambda$ will not by 
itself cause complete phase loss, i.e. we are in the regime of phase wandering.
A TLS with $E>T_\Lambda$ causes a phase jump. Let us first assume 
that 
\begin{equation} 
  \label{eq:ineq102} 
\Lambda_0 \ll T_\Lambda \ll T\, . 
\end{equation} 
 Using the distribution (\ref{dm}) and 
returning to dimensional time one obtains 
\begin{equation} \label{eq:aux1} 
\bar{\Gamma}_1 (t) \approx 
  \frac{t}{\tau_i}\,\frac{\Lambda^2_0}{T_\Lambda E^*} + 
  \frac{t}{\tau_i} \, \left(\frac{\Lambda_0 
  t}{\hbar}\right)^2 \frac{T_\Lambda}{E^*}\, 
  . 
\end{equation} 
Now, let us define $\tau_\varphi^{(1)}$ and $T_\Lambda$ to make both
  contributions to $\bar{\Gamma}_1 (\tau_\varphi^{(1)})$ equal to 1. This 
definition of $T_\Lambda$ is consistent with that given above within the 
accuracy of the approximation because the two terms are the expansions 
in large and small values of $E \tau_\varphi/\hbar$. The point where 
both terms becomes of the order 1 should then correspond to the crossover 
point $E \tau_\varphi/\hbar=1$. This is easily checked from the formulas 
below.  In this way we get
\begin{equation} \label{eq:aux2} 
  \tau_\varphi^{(1)}=\tau_\Lambda (T_\Lambda/\Lambda_0)\, , 
  \end{equation} 
where 
\begin{equation}
T_\Lambda=(\hbar \Lambda_0/\tau_\Lambda)^{1/2} \, \quad
\tau_\Lambda=\tau_i(E^*/\Lambda_0)\, .
\end{equation} 
The time $\tau_\Lambda$ is due to the dynamic defects with symmetric 
potentials and 
energy splitting equal to $\Lambda_0$. At $T\le T_\Lambda$ for all 
energies the inequality $E\tau_\varphi /\hbar \ll 1$ is met, and 
only the second item in Eq,~(\ref{eq:aux1}) is important.  One 
should replace $T_\Lambda$ by $T$ in this expression to obtain 
$\tau_\varphi^{(1)}= \tau_\Lambda 
(T_\Lambda/\Lambda_0)(T_\Lambda/T)^{1/3}$. In contrast, if 
$T_\Lambda \le \Lambda_0$ then only the first item in 
Eq,~(\ref{eq:aux1}) is important. In this case $ \tau_\varphi^{(1)}= 
\tau_\Lambda$. 
 
  The result can be summarized  as 
\begin{equation} 
  \label{eq:tauphi01} 
  \frac{1}{\tau_\varphi^{(1)}} \approx \frac{1}{ 
    \tau_\Lambda}\frac{\Lambda_0}{T_\Lambda} 
\left\{\begin{array}{ll} 
\min\{(T/T_\Lambda)^{1/3}, 1\}\, , & T_\Lambda \gg \Lambda_0 \, , \\ 
T_\Lambda/\Lambda_0 \, & T_\Lambda \ll \Lambda_0 \, . \end{array}\right. 
\end{equation}

\paragraph {Contribution of relaxation processes.} 
 
Since only $E\lesssim T$ are important, for estimates  one can 
assume $E\coth E/2T \approx 2T$. Thus, $\gamma_0 \approx 2\chi 
T$ becomes $E$-independent. For the same reason 
$\tau_3^{-1}$ can be approximated as 
$\tau_i^{-1}(\Delta/E)$. The following calculation 
depends on the relationship between $\gamma$ and $\tau_\varphi$. 
At $\gamma \tau_\varphi  \ll 1$ one can expand the expression for 
$\Gamma_3$ as 
\begin{equation} 
\label{eq:gamma201} 
\Gamma_3(\eta,E,\Lambda) \approx \eta^2 \gamma \tau^2 /\tau_3 
=\gamma_0 (\eta \tau)^2 
\tau_i^{-1}\Delta^2 \Lambda^2/E^4\, . 
\end{equation} 
while at $\gamma \tau_\varphi  \gg 1$ one has 
\begin{equation} 
  \label{eq:gamma202} 
\Gamma_3 =\eta 
(\tau/\tau_3) \propto (\Delta/E)^2\, . 
\end{equation} 
To estimate $\bar{\Gamma}_3 (\eta)$ let us introduce the energy 
splitting $E_\chi$ at which $\gamma \tau_\varphi=1$. The meaning of this is 
that a TLS with $E<E_\chi$ will probably jump during the trajectory traversal 
time $\tau_\varphi$ (it is ``fast moving''), whereas a TLS with $E>E_\chi$ 
will have a low probability to jump in the same time (it is ``slow moving'').
First we assume that 
\begin{equation} 
  \label{eq:ineq301} 
\Lambda_0 \ll E_\chi \ll T\, . 
\end{equation} 
 Using the distribution (\ref{dm}) and 
returning to dimensional time one obtains 
\begin{equation} \label{eq:aux} 
\bar{\Gamma}_3 (t) \approx \frac{t}{\tau_i}\frac{E_\chi}{E^*} + 
  \frac{\chi T t^2}{\hbar \tau_i}\frac{\Lambda_0^2}{E_\chi E^*}\, 
  . 
\end{equation} 
Now, let us define $\tau_\varphi$ and $E_\chi$ to make both contributions 
to $\bar{\Gamma}_3$ equal to 1. This procedure indicates 
that the defects with $E=E_\chi$ are those which experience a hop 
during the typical Cooperon trajectory traversal time. One obtains 
  \begin{equation} 
    \label{eq:tauphi301} 
    1/\tau_\varphi^{(3)} = 
    \tau_i^{-1}(E_\chi/E^*)\, , \quad E_\chi=\Lambda_0 \left(\chi T 
    \tau_\Lambda/\hbar\right)^{1/3} \, . 
  \end{equation} 
Introducing the characteristic temperature 
$T_\alpha=\hbar/\chi \tau_\Lambda$ 
at which $E_\chi=\Lambda_0$ one can express the dephasing 
rate as 
\begin{equation} 
    \label{eq:tauphi304} 
    1/\tau_\varphi^{(3)}= 
    \tau_{\Lambda}^{-1}(T/T_\alpha)^{1/3} \, . 
  \end{equation} 
Another important characteristic energy is the temperature $T_\beta$ 
at which $E_\chi=T$, 
$T_\beta=\Lambda_0^{3/2}/T_\alpha^{1/2}$  The 
ratio $T_\alpha/T_\beta=(\chi \Lambda_0 \tau_\Lambda/\hbar)^{3/2}$ 
can be arbitrary. 

The meaning of $T_\alpha$ and $T_\beta$ can be understood as follows. 
Imagine starting at some large temperature where $T\gg E_\chi \gg \Lambda_0$. 
As we lower the temperature $E_\chi$ is also decreasing, but it decreases at
a slower rate than $T$ ($E_\chi\sim T^{1/3}$). At the temperature $T_\alpha$,
$E_\chi=\Lambda_0$. Since $\Lambda_0$ is the lower cutoff for $E$, 
if $T<T_\alpha$ \, all defects are slow-moving because no defects exist 
with sufficiently low splitting. Alternatively, since $T$ is decreasing 
faster than $E_\chi$, $T$ will overtake $E_\chi$ at the temperature $T_\beta$.
For $T<T_\beta$ all defects are fast, because the slow ones are frozen out.
Which temperature is reached first depends
on the specific values of the parameters (since the ratio $T_\alpha/T_\beta$
is arbitrary). Also, if $T_\alpha>T_\beta$ then $E_\chi<\Lambda_0$ for all 
$T<T_\alpha$, so in particular $T_\beta<\Lambda_0$ and is thus unimportant.
Similarly, if $T_\alpha<T_\beta$ then $T_\alpha<\Lambda_0$.
 
The result (\ref{eq:tauphi301}) is valid if $T \gg E_\chi \gg 
\Lambda_0$, or at $T \gg T_\alpha, T_\beta$. At  $E_\chi \gg T \gg 
\Lambda_0$, or $T_\beta \gg T \gg \Lambda_0$, only 
the first item 
in Eq.~(\ref{eq:aux}) exists and the quantity $E_\chi$ should be 
replaced by $T$. As  a result, 
\begin{equation} 
    \label{eq:aux3} 
1/\tau_\varphi^{(3)} \approx 
\tau_\Lambda^{-1} (T/\Lambda_0)\, . 
\end{equation} 
Since $\Lambda_0=T_\beta^{2/3}T_\alpha^{1/3}$ the results 
(\ref{eq:tauphi304}) and 
(\ref{eq:aux3}) match at $T=T_\beta$ 
This temperature region exists only 
if $T_\beta > T_\alpha$. 
At $T \lesssim \Lambda_0$ the 
dephasing rate strongly decreases with the temperature decrease. 
 
\begin{figure}[h] 
\centerline{
\includegraphics[width=7cm]{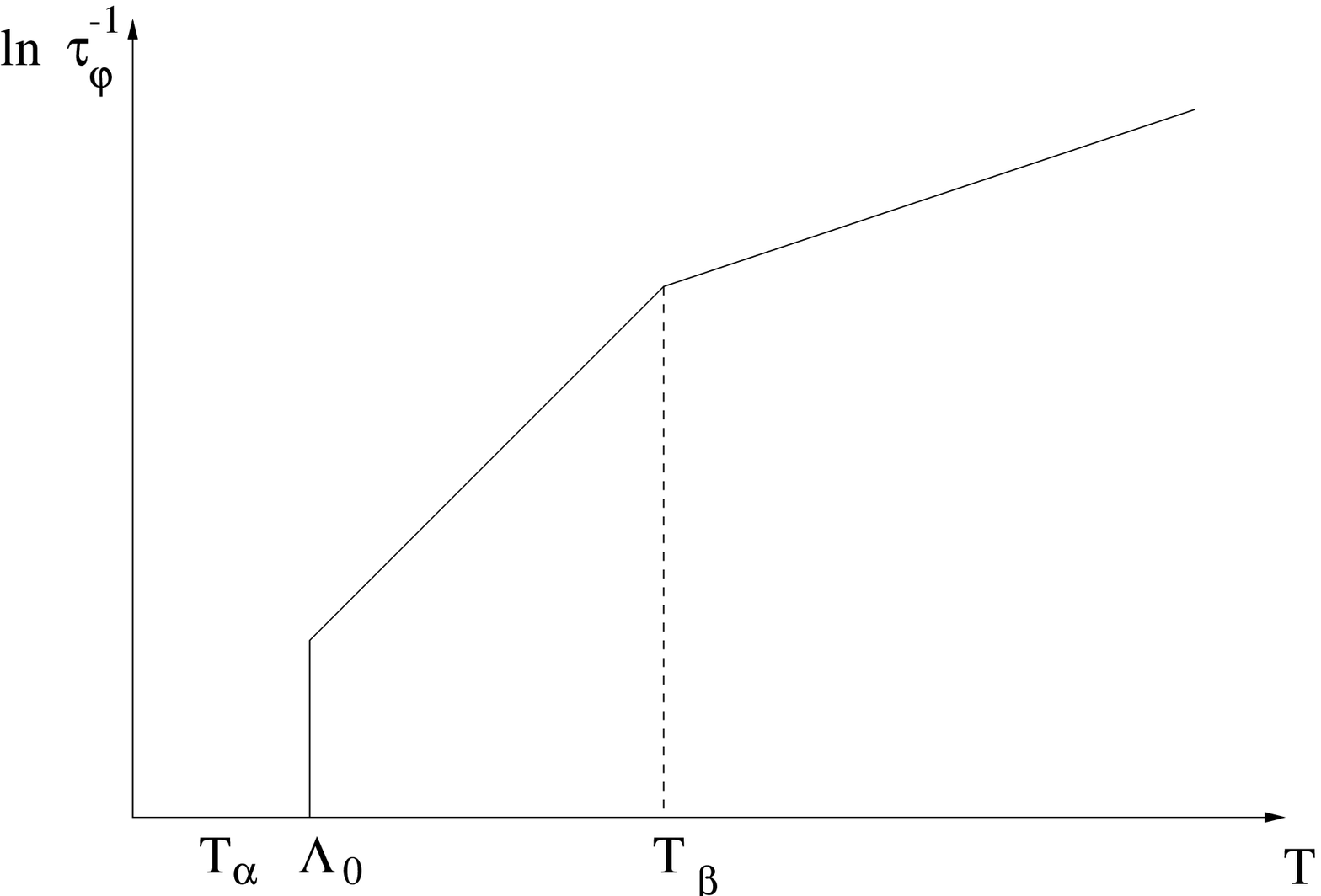}
} 
\caption{Schematic picture of $\ln\frac{1}{\tau_\varphi}$ 
as function of temperature for $T_\alpha<T_\beta$. \label{f-r1}} 
\end{figure}

\begin{figure}[h] 
\centerline{
\includegraphics[width=7cm]{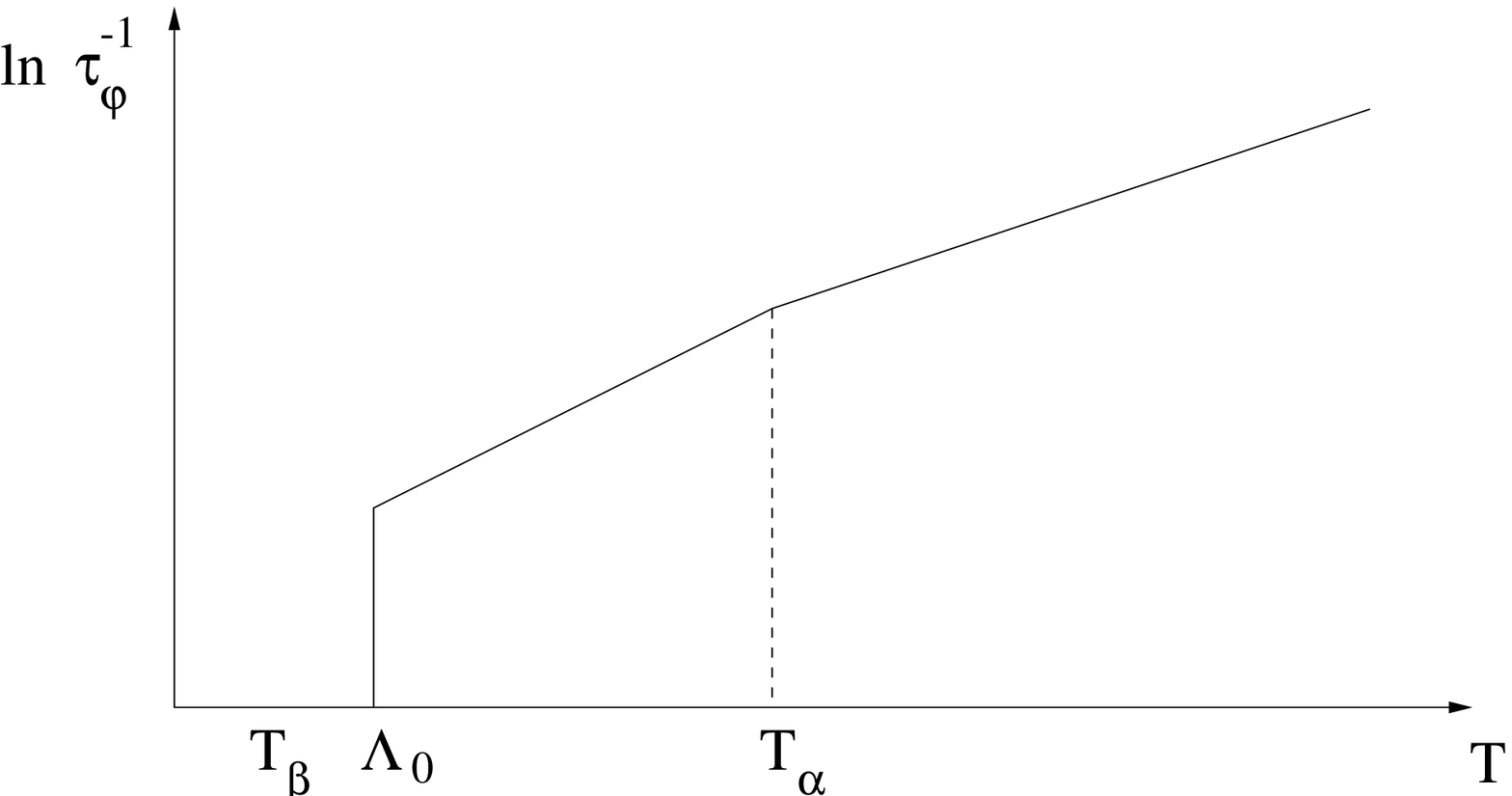}
} 
\caption{Schematic picture of $\ln\frac{1}{\tau_\varphi}$ 
as function of temperature for $T_\alpha>T_\beta$. \label{f-r2}} 
\end{figure}

If $T_\alpha \gg T \gg T_\beta$
the relaxation is slow for all the dynamic defects 
and the second item in Eq.~(\ref{eq:aux}) is important. However, in this 
case one has to replace $E_\chi \to \Lambda_0$ in its estimate. As a 
result, 
\begin{equation} \label{eq:tauphi303} 
  1/\tau_\varphi^{(3)}= 
  \tau_{\Lambda}^{-1} \left(T/T_\alpha\right)^{1/2} \, . 
\end{equation}
The temperature dependence of $\tau_\varphi$ 
is sketched in Fig. \ref{f-r1} for $T_\alpha<T_\beta$
and in  Fig. \ref{f-r2} for $T_\alpha>T_\beta$.

\paragraph{Comparison} 
 
Now we are in a position to compare the contributions to the dephasing 
rate. Both contributions to $1/\tau_\varphi$ are parametrized by the 
quantity $1/\tau_\Lambda$, the relative contributions being dependent 
on the temperature. The relative resonant contribution crosses over 
from $(\Lambda_0/T_\Lambda)(T/T_\Lambda)^{1/3}$ to $\Lambda_0/T_\Lambda$ 
at $T=T_\Lambda$. The 
relative relaxation contribution crosses over from $T/\Lambda_0$ to 
$(T/T_\alpha)^{1/3}$ at $T=T_\beta$ if $T_\beta \gtrsim T_\alpha$. In 
the opposite case is crosses over to $(T/T_\alpha)^{1/2}$ at 
$T=\Lambda_0$ and then to  $(T/T_\alpha)^{1/3}$ 
at $T=T_\alpha$. 
 
We conclude that at $T \ge T_\alpha, T_\beta$ the relaxation 
contribution dominates, and the dephasing rate is 
\begin{equation} \label{drate} 
\tau_\varphi^{-1} = \tau_{\Lambda}^{-1}[(T/T_\alpha)^{1/3} + \zeta] 
\end{equation} 
 where $\zeta$ is a 
constant of the order 1 originating from the resonant 
contribution. 
 
At low temperatures both contributions can be important, their 
interplay depending on the relationship between the temperature 
$T$ and the characteristic energies $\Lambda_0$, $T_\Lambda$, 
$T_\alpha$  and $T_\beta$. For both mechanisms the dephasing rate 
vanishes as $T \to 0$ and there is is a region in which the 
dephasing rate is proportional to $T^{1/3}$ .

\paragraph{Averaging over the tunneling matrix element.} 
 
In our considerations we have assumed that ${\cal P}_{TM} \propto 
\delta (\Lambda - \Lambda_0)$, i.~e. that the tunneling  
matrix element $\Lambda$ is given. Note, however, that due to  
disorder the barrier parameters are also  
scattered. To discuss role of such a scatter let us assume that the 
overlap integral $\Lambda$ is given by the expression $\Lambda= 
( \hbar \omega_0/\pi)\,  e^{-\lambda}$ where $\omega_0$ is some 
attempt frequency  while 
the barrier   
strength $\lambda$ is distributed according to Gaussian law around 
some central value $\lambda_0=\ln \hbar \omega_0/\pi \Lambda_0 \gg 1$, 
\begin{equation}\label{Gs} 
 {\cal P}_\lambda = {\cal N}\,  e^{-(\lambda - 
\lambda_0)^2/2\bar{\lambda}^2}\, . 
\end{equation} 
Here ${\cal N}$ is a proper normalization factor which at 
$\bar{\lambda} \ll \lambda_0$ is equal to $(2\pi \bar{\lambda}^2)^{-1/2}$. 
As we have seen, for most interesting regimes it is the quantity 
 $\Lambda^2$ that has to be averaged. The results then can be 
 expressed as 
$$\frac{\overline{\Lambda^2}}{\Lambda_0^2}= {\cal N}\int_{-\lambda_0}^\infty 
d\xi \,e^{-2\xi -\xi^2/2\bar{\lambda}^2}  \approx 
e^{2\bar{\lambda}^2} \quad \text {at} \quad  \lambda_0 \gg
\bar{\lambda} \ \, .$$    
 As it is seen, the 
only effect of the scatter in $\lambda$  corresponds to 
renormalization of the tunneling matrix element by a constant factor $e^{\bar 
  \lambda^2}$. Since this factor is 
temperature-independent, 
it does not affect qualitatively the picture obtained with an 
assumption of a fixed value of $\Lambda$. One also notes that 
the averaging procedure practically cuts out a contribution of the 
region $\lambda > \lambda_0$ since at this region the tunneling 
matrix element exponentially decays with $\lambda$. Thus the 
picture is not sensitive to this region, and any distribution of 
$\lambda$ with a lower cut-off (even allowing a Gaussian smearing 
of this cut-off) does not change the considerations made in an 
assumption of a fixed value of $\Lambda = \Lambda_0$.

 

\section{Estimates and discussion} \label{estimates}  
 
To make estimates we rewrite the expression (\ref{eq:inr}) for 
$\tau_i$ in the form  
\begin{equation}\label{taud} 
\tau_i^{-1} = \sigma_{\text{in}} v_F n_d\, , \quad 
\sigma_{\text{in}} \equiv 
\sigma_{e}^{\text{d}} \,|V^{-}/V^{+}|^2 
\end{equation} 
where $\sigma_{e}^{\text{d}} $ is the cross-section of elastic electron 
scattering by a dynamic defect. Correspondingly, the key parameter of our 
theory, $\tau_{\Lambda}$, is given as 
\begin{equation}\label{tauL} 
\tau_{\Lambda}^{-1} = \Lambda_0 P_d \sigma_{\text{in}} v_F 
\end{equation} 
where $P_d = n_d/E^{*}$ is the density of states of the  
dynamic defects. 
 
The density of states $P_d$ can be, in principle, estimated for a 
given material on the basis of point contact measurements. Namely, 
metallic point contacts are known to exhibit, first, telegraph 
resistance noise\cite{Cornell1} and, second, zero-bias anomalies; 
\cite{Cornell2} both effects are expected to be associated with 
the dynamic defects.\cite{Cornell1,Cornell2,kr}  
 
Although we appreciate that the material preparation procedure can 
significantly affect the defect system, we believe that 
such 
experiments can provide more or less reasonable estimates  
for $P_d$.  The telegraph noise studies\cite{Cornell1} for 
a Co nanoconstriction with a size of $\sim 10$ nm  
revealed the presence of about several dynamic defects  at energies less 
than 10 mV. This would give us the value $P_d \sim  
(3-5)\times 10^{32}$ erg$^{-1}$cm$^{-1}$. However, the telegraph noise is 
related to TLS with rather slow relaxation rates ($\lesssim 10^3$ 
 s$^{-1}$) while we are interested in the defects with switching times 
of the order of $10^{-9}$ s. Consequently,  these estimates most probably 
significantly \emph{underestimate} 
$P_d$. What is more instructive, the magnitude of the resistance noise 
revealed rather large defect asymmetry corresponding to the 
estimate $\sigma_{\text{in}} \sim \sigma_{e}^{\text{d}}  \sim 
10^{-15}$~cm$^2$.  
 
We believe that the zero bias anomalies can give more reliable 
information concerning $P_d$. The magnitude of these anomalies 
for Co nanoconstrictions\cite{Cornell2} of the same type as 
mentioned above corresponds to a presence of several tens of TLS 
at the energy region about 1~meV.\cite{Cornell2,kr} 
Correspondingly, one obtains $P_d \sim (3-5)\times 
10^{34}$~erg$^{-1}$cm$^{-3}$.  
 
Based on these estimates and taking $P_d \approx 10^{34}$ 
erg$^{-1}$cm$^{-3}$, $\sigma_{\text{in}} \approx 10^{-15}$ cm$^2$,  
$v_F \approx 10^8$ cm/s, and $\Lambda_0 \approx 10$ mK we obtain 
$\tau_{\Lambda}  
\approx 10^{-9}$ s. 
Equations (\ref{taud}) and 
(\ref{tauL})  
yield $T_{\Lambda} 
\simeq \Lambda_0$. Thus at temperatures larger than $T_\Lambda \approx 
\Lambda_0 \approx  10$ mK  one expects, according to Eq.~(\ref{eq:tauphi01}), 
temperature-independent contribution of resonant processes. 
 
For the relaxation channel, one obtains $T_{\alpha} \approx T_{\beta} 
\approx 10$ mK. Consequently, at $T \gtrsim T_\alpha \approx T_\Lambda 
\approx 10$ mK one expects that 
dephasing rate obeys  Eq.~(\ref{drate}) 
with $\tau_\Lambda \approx 10^{-9}$~s.

Now let us check if our assumption $\Lambda_0 \approx 10$ mK is  realistic.  
We will exploit a crude estimate 
\begin{equation} 
\Lambda_0 \simeq \frac{\hbar \omega_0}{\pi}  \, \exp \left(- 
  \frac{2}{\hbar}  \int_0^{a }  
d r\,  \sqrt{2 M U(r)}\right) 
\end{equation} 
where $U(r)$ is a potential relief between the two stable defect 
positions separated by a distance $a $, 
and M is the defect mass. Taking as an example 
$U(r) = (U_0/2)\left[1 - \cos(2\pi r/a )\right]$ one obtains for the 
exponent 
$ (2a /\pi \hbar) \sqrt{2 U_0 M} $. 
Taking for a light defect $\omega_0 \approx 10^{14}$ s$^{-1}$ and 
assuming $a  \approx 10^{-8}$ cm, $U_0 \approx 0.2$ eV one 
estimates that the value $\Lambda = 10$ mK is achievable for $M 
\approx 2 \times 10^{-23}$~g which corresponds to atomic weight $\approx 
10$.

Summarizing our estimates, we can conclude that for realistic 
parameters of the dynamic defects one can indeed expect a slow temperature 
dependence of the dephasing rate given by Eq.~(\ref{drate}) crossing 
over to a rapid decrease   
at low temperatures. The crossover temperature, as well as the 
behavior below that temperature, 
 depends on 
the distribution of $\Lambda$. For a delta-like distribution 
(\ref{dm}) the TLS spectrum has a gap 
of $\Lambda_0$. 
Thus the TLS contribution to dephasing rate is exponentially frozen out  
at 
for  $T < \Lambda_0$, and we are left with the ``standard" 
mechanisms like electron-electron scattering. However for the 
Gaussian distribution (\ref{Gs}) with $\bar {\lambda} \gg 1$  the 
situation is 
different. In this case the cut-off temperature is given 
by the renormalized tunneling coupling, $ \Lambda_0 e^{\bar \lambda^2} $ 
while for lower temperatures one deals with rather flat 
distribution of $\lambda$ within the region $\lambda \le \lambda_0 + 
\bar{\lambda}$. Correspondingly, at these temperatures one deals with 
a glass-like TLS distribution for which $\tau_{\varphi} \propto 
T^{-1}$.

Although some papers, e. ~g. Refs.~\onlinecite{Ahn,Aleiner}, stated
that to explain the dephasing saturation  by a TLS
contribution one would need an unreasonably large concentration of the
TLS, this conclusion was mainly based on the ``glassy" model of the TLS
while we exploited the tunneling state model of
Refs.~\onlinecite{kr,Phillips1}.  In general, to obtain independent
information concerning the TLS concentration based on ``bulk"
measurements like acoustic measurements or heat capacity measurements
in conductors is rather difficult due to a presence of electronic
contributions. In particular, the value $P_d \sim 10^{34}$ erg$^{-1}$
cm$^{-3}$ exploited above is still less than the electron density of
states ($\sim 10^{35}$ erg$^{-1}$cm$^{-3}$ for Co) and so the TLS are
not expected to affect significantly the properties of the material,
like heat capacity.
 
Now we would like to compare our results with the calculations given
in  Refs.~\onlinecite{ifs,RDZ,Ahn} where a similar problem was considered.
 The authors of
Ref.~\onlinecite{Imry} gave a semi-phenomenological treatment of the
problem. They exploited the TLS distribution typical for the standard
glassy TLS model, but with the upper cut-off $\Lambda_{0,\max}$ for
the tunneling matrix element. For the resulting dephasing time they
reported a proportionality of $\tau_{\varphi}^{-1}$ to
$\Lambda_{0,\max}/E^{*}$ (in our notations) in the limit
$\tau_{\varphi}\Lambda_{0,\max} > \hbar$. One notes that such a
proportionality is in agreement with the second line of our
Eq.~(\ref{eq:tauphi01}) although the total expression for
$\tau_{\varphi}$ was some different from ours. Furthermore, the
estimate for the opposite limiting case was completely different from
our Eq.~(\ref{eq:tauphi01}). Since Eq.~(\ref{eq:tauphi01}) corresponds
to the ``resonant" or ``inelastic" channel we conclude that the authors
of Ref.~\onlinecite{Imry} accounted for only these inelastic processes
of electron dephasing. The ``elastic", or $\sigma_3$,  channel (which,
as we have seen, 
can dominate with respect to the ``inelastic" one) seems to stay beyond
the quantitative results of Ref.~\onlinecite{Imry}.
 
In Ref.~\onlinecite{RDZ} the dephasing due to dynamical defects was
treated within the framework of the two-channel Kondo-model. We
believe that this model is not relevant to the metallic samples we are
interested in, see Ref.~\onlinecite{kondo}.

Then, the dephasing by TLS was also considered in the recent paper 
Ref.~\onlinecite{Ahn} where the saturation behavior of $\tau_\varphi$
in quantum dots for the TLS distribution with fixed $\Lambda_0$ was
claimed. As it follows from the derivation,\cite{Ahn} only the
transitions between TLS states due to interaction with the electron
forming the interference loop are taken into account. At the same
time, the transitions due to other electrons forming a thermal bath
(second mechanism of dephasing) are ignored. Hence, only the
$\sigma_1$ channel is taken into account,and the result is
similar to the second item in our Eq.~(\ref{eq:tauphi301}). However,
the main contribution arising from the $\sigma_3$ channel is omitted. 
 
It is worthwhile to mention that similar ideas were used to explain
the  magnetoresistance of polymers.~\cite{Polymer} Polymer 
systems exhibiting rather large fraction of free volume are 
expected to form readily mobile and metastable defects of different types.

\section{Conclusions} \label{conclusions} 
 
To conclude, we have shown that the dynamic defects can be 
responsible for the slowing down of the temperature dependence of the
dephasing rate at low temperatures. 

There are two mechanisms of dephasing. The first one corresponds 
to direct inelastic scattering of electrons by the defects,
while the second one is due to violation of the time reversal symmetry
caused by fluctuations of the scattering potential.  
 The first mechanism  can indeed lead to the saturation, while the 
second one still contains a temperature dependence although a weak 
one.  

However, when $T \lesssim \Lambda_0$ the dephasing rate rapidly tends to 0.

\begin{acknowledgments} 
We thank Y. Imry for discussions.
This research is partly supported by the Norwegian Research Council 
and partly by the  
US DOE Office of Science under contract No. W-31-109-ENG-38.  
\end{acknowledgments}

\appendix 
\section{Calculation of the relaxation rates} 
The relaxation rate is determined as an imaginary part of the 
analytically-continued Matsubara  
self-energy. In general, we have 3 contribution to the self energy due 
to three different types of  
the electron-TLS interaction. Since for a short-range scattering 
potential the interaction vertexes  
do not have an internal structure, for each bosonic propagator ${\cal 
  D}_i$ one obtains:  
\begin{eqnarray} 
\Sigma^M_i(\varepsilon_k)&=&\rho g_i^2 \int d\xi_p\, F^M_i(\varepsilon_k, \xi_p)\, , \nonumber \\ 
F^M_i(\varepsilon_k,\xi_p)&=&T \sum_s{\cal D}^M_i (\omega_s)G(\varepsilon_k - \omega_s, \xi_p)\, . 
\end{eqnarray} 
Here $\rho$ is the electron density of states, $\omega_s=2\pi sT$, $\varepsilon_k=2\pi(k+1)T$, $\xi_p =p^2/2m - \mu$ 
and $g_i$ is the proper coupling constant determined by the Hamiltonian (\ref{dh}). 
The analytical continuation is performed in a usual way. Since there are two cuts in the complex 
$\omega$-plane, at $\Im\, \omega=0$ and $\Im\, (\varepsilon - \omega)=0$ for each $i$ we get 
$F^R(\varepsilon_k,\xi_p)=F_1+F_2$ where 
\begin{eqnarray} 
F_1&=& \int_{-\infty+\varepsilon_k}^{\infty +\varepsilon_k 
}\!\! \!\frac{ N(\omega){\cal D}^R(\omega)\, d\omega}{2\pi i} 
\left[G_A(\varepsilon_k-\omega) -G_R(\varepsilon_k-\omega)\right]\nonumber\\ 
F_2&=&\int_{-\infty}^{\infty}\frac{ N(\omega) G_R(\omega)\, d\omega}{2\pi i}\left[ 
{\cal D}^R(\omega)-{\cal D}^A(\omega) 
\right]\, . 
\end{eqnarray} 
For brevity we omit the arguments $\xi_p$ pf the electron Green's functions. 
 Now we replace the integration variable in the expression for $\Sigma_1$ as $\omega \to \omega-\varepsilon_k$ 
 and then combine two integrals. Taking in account that $N(\omega +i\pi T)=-n(\omega)$ 
 where $n(\omega)=\left(e^{\omega/T} +1\right)^{-1}$, $\Im \, G_R (\varepsilon, \xi_p)=\pi \delta(\varepsilon - \xi_p)$ 
  and making a straightforward algebra we obtain 
\begin{eqnarray*} 
\Im \, F^R(\varepsilon,\xi_p)&=&-\int_{-\infty}^{\infty} 
d \omega\,\left[\coth\left(\frac{\omega}{2T}\right)+\tanh\left(\frac{\varepsilon -\omega}{2T}\right)\right] \\ 
%
&&\times \,  \Im \left[ {\cal D}^R (\omega)\right]\, 
\delta(\varepsilon -\omega - \xi_p)\, .  
\end{eqnarray*} 
Performing trivial  integration over $\xi_p$ and taking into account the $1/2\tau = - \Im  \Sigma_R$ we finally obtain 
\begin{eqnarray} 
\tau_i^{-1}(\varepsilon)&=&2\pi\rho g_i^2\, {\cal G}_i(\varepsilon)\, , 
\label {tau00} \\  
{\cal G}_i (\varepsilon)&=&\int_{-\infty}^\infty \! \!d \omega\, 
N(\omega)n(\varepsilon - \omega) n^{-1}  
(\varepsilon) \, \Im \left[ {\cal D}_i^R (\omega)\right]\, . \label{F00} 
\end{eqnarray} 
Using Eq.~(\ref{D1}) we obtain 
\begin{equation} 
\label{tau1_1} 
{\cal G}_1=n(\varepsilon + E)\, n(\varepsilon -E)\, n^{-2}(\varepsilon)\, . 
\end{equation} 
At $E \gg T$ it is proportional to $e^{-E/T}$ at any finite $\varepsilon$. 
 
While calculating ${\cal G}_3$ one can expand $N(\omega) \approx 
T/\omega$, $n(\varepsilon-\omega)/n(\varepsilon) \approx 1$. After 
that 
\begin{equation} 
\label{tau1_3} 
{\cal G}_3=\cosh^{-2}(E/2T) \, . 
\end{equation} 
It can be easily shown that ${\cal G}_0=1$.

\section{Mapping to a random-telegraph-noise model} \label{rtn} 
 
Consider an electron trajectory with the total traversal time 
$t_0$ which contains $N$ dynamic defects able to hop between two 
sites. 
They are rather rare, so a typical neighbor of any active 
dynamic defect is a {\sl static} one . The total length of the trajectory is 
$$ 
{\cal L}_0=\tau/v_F=\sum_{s=1}^M \left|\bR_{s+1}^{(0)}-\bR_s^{(0)}\right| 
$$ 
where $M$ is the total number of defects, $M \gg N$.  
 
Let us parameterize the electron motion along the trajectory by 
time $t$ and allow some of the defects (labeled by $j$) to make
transitions between their states. 
For those defects, 
$$ 
\bR_{j}(t) =\bR_j^{(0)}+ \bu_j(t)\, . 
$$ 
The length of distorted trajectory $\cL^+$ traversed in the {\sl 
positive} direction is 
\begin{eqnarray*} 
\cL^+&=&\sum_{s=1}^M 
\left|\bR_{s+1}^{(0)}+\bu_{s+1}(t_{s+1})-\bR_s^{(0)}- 
\bu_{s}(t_{s})\right| \nonumber \\ 
&=& \cL^{(0)}+ v^{-1}\sum_{j=1}^N {\tilde \bv}_j \cdot 
\bu_j(t_j)\, . 
\end{eqnarray*} 
Here $ {\tilde \bv}_j \equiv (\bv_{j-1}-\bv_{j})$ is the change in 
the electron velocity due to scattering by $j$th EF. 
 
Now we can specify the displacement of $j$th EF as 
$$ 
\bu_j(t)\equiv \ba_j\, \xi_j(t) 
$$ 
where $\xi_j(t)$ is a {\em random telegraph process} (RTP), i.e. a function 
 switching between the values $\pm 1$ at random times and having the 
correlation function 
$$ 
\langle\xi_j(t)\xi_k(t') \rangle=\delta_{jk}e^{-2\gamma_j|t-t'|}\, . 
$$ 
 Then the 
time-dependent contribution to the length 
is $$ (\delta \cL)_j (t)=l_j\, \xi_j(t)\, , \quad l_j \equiv 
(\bv_j\cdot \ba_j)/v \, .$$ 
For a given defect $j$, 
the phase difference is just 
$$ 
(\delta \Phi)_j(\tau)=(p_Fl_j/\hbar)\, [\xi(t_j) - \xi(t_0-t_j)]\, . 
$$ 
Let us split the calculation of the average $\langle e^{i\, \delta 
\Phi} \rangle$ in two steps. First let us calculate the average 
over different realizations of a given RTP, 
$$k(\tau)=\left \langle e^{iJ\, [\xi(t)-\xi(\tau 
-t)]}\right \rangle_{RTP}\, , \quad J \equiv p_Fl/ \hbar\, . $$ 
This sum can be calculated using the generation function (for $t_\beta
>t_\alpha$)  
\begin{eqnarray*} 
K(x,y)&=&\left \langle e^{-ix\xi(t_\alpha)-iy\xi(t_\beta)}\right\rangle 
\nonumber \\ 
&=& e^{-\gamma(t_\beta-t_\alpha)}\left[\cos (x+y) \cosh
  \gamma(t_\beta-t_\alpha)\right.  
\nonumber \\ && \left. 
+ \cos 
(x-y) \sinh  \gamma(t_\beta-t_\alpha)\right]\, . 
\end{eqnarray*} 
Substituting $x=-y=J$ and considering arbitrary times we obtain 
$$ 
k(t,t_0) 
=2\cos^2J+ 2\sin^2J\, e^{-2\gamma |t_0 -2t|}\, . 
$$ 
We observe that the function depends explicitly on the position of the 
scatterer along the trajectory, that is natural. It does not contain complete 
destruction of the interference -- it appears only after averaging 
over different dynamic defects. 
 
The average over different dynamic defects will be performed using the
Holtsmark procedure (see, e.~g., Ref.~\onlinecite{Chandrasekhar} for a
review) according to which 
$$
\langle e^{i (\Delta \Phi)} 
\rangle_{d} = e^{-W(t_0)}\, , \quad W(t_0) \equiv n_{\text{eff}}V_c
\kappa (t_0)$$ 
Here $n_{\text{eff}}$ is the concentration of ``active'' defects, $V_c$
is the ``contact volume'', while 
$$W(t_0) = \langle 1-k(t,t_0) \rangle_{d}= \left \langle  2\sin^2 J\,
 \eta(t_0-2t) \right \rangle_{d}$$
where $\eta(t) \equiv 1-e^{-2 \gamma |t|}$. 
The contact volume is estimated as $V_c = \sigma v_F t_0$, where
$\sigma $ is the  scattering cross section. 

Let us for simplicity assume that the hopping distances $a_j$ of the defects 
are the same. We have also to assume that the phase changes due to 
individual hops to be small to keep the treatment consistent. Assuming 
$J \ll 1$ we easily average over the directions of hops to get 
$\overline{J^2}=(4\pi^2/3)(p_Fa/\hbar)^2$. Now let us average over the
positions of  
the defects along the trajectories. This is done as 
$$
\kappa(t_0)= 2 \overline{J^2} \int_0^{t_0/2} \! \!\eta (t_0-t) \,
\frac{dt}{t_0} 
=\frac{\overline{J^2}}{2\gamma} \, \left(2\gamma -1+e^{-2 \gamma
    t_0} \right) \, .
$$
We observe that $\kappa (t_0) \sim \overline{J^2}  \min\{\gamma t_0,1\}$. 
Collecting the factors, we obtain
$$W(t_0) \approx (t/\tau_3) \min\{\gamma t_0,1\}$$
where 
$$\tau_3^{-1} \approx   (4\pi^2 n_{\text{eff}}/3)(p_Fa/\hbar)^2 \sigma
v_F\, .$$ 
The concentration $n_{\text{eff}}$ depends on the distribution of the
TLS parameters. After evaluating it in a proper way one recovers the
results of Eq.~(\ref{eq:aux1}).


\begin{thebibliography}{99} 
 
\bibitem{heiblum1} 
 E. Buks, R. Schuster, M. Heiblum, D. Mahalu, V. Umansky and H. Shtrikman, 
\prl {\bf 77}, 4664 (1996). 
\bibitem{heiblum2} 
A. Yacoby, M. Heiblum, D. Mahalu, Hadas Shtrikman, \prl {\bf 74}, 047 (1995). 
\bibitem{mw} P. Mohanty, E. M. Q. Jariwala, and R. A. Webb, \prl {\bf 
78}, 3366 (1997). 
\bibitem{gz} D. S. Golubev and A. D. Zaikin, \prl, {\bf 81}, 1074 (1998). 
\bibitem{aag} I. L. Aleiner,  B. L. Altshuler, and M. E. Gershenson, 
\prl {\bf 82},3190 (1999). 
\bibitem{mis} P. Mello, Y. Imry, B. Shapiro, \prb {\bf 61}, 
16570 (2000). 
\bibitem{ifs} Y. Imry, H. Fukyama, and P. Schwab, Europhys. Letters, 
{\bf 47}, 608 (1999). 
\bibitem{RDZ} A. Zawadowski, Jan von Delft and D. C. Ralph, \prl {\bf 83},
2632 (1999).
\bibitem{Ahn} Kagn-Hun Ahn, P. Mohanty, \prb {\bf 63}, 
195301 (2001) 
\bibitem{ahvp} P. W. Anderson, B. I. Halperin, and C. M. Varma, 
  Philos. Mag. {\bf 25}, 1 (1972). 
\bibitem{Black} J. L. Black, in \emph{Glassy Metals 1}, edited by 
  H. J. G\"unterodt and H.~Beck (Springer, Berlin 1981). 
\bibitem{Stern} A. Stern, Y. Aharonov, and Y. Imry, \prb {\bf 41}, 3436 (1990);
    Y. Imry, \emph{Introduction to Mesoscopic Physics}, 
   (Oxford University Press, 1997).
\bibitem{Schwab} P. Schwab,  Eur. Phys. J. B {\bf 18}, 189 (2000)
\bibitem{aalk} B. L. Altshuler, A. G. Aronov, and D. E. Khmelnitskii, 
  J.\ Phys. C {\bf 15}, 7367 (1982); see also  
B. L. Altshuler, A. G. Aronov, A. I. Larkin, 
D. E. Khmelnitskii, Sov. Phys. JETP {\bf 54}, 411 (1981). 
[Zh. Eksp. Teor. Fiz. {\bf 81}, 768 (1981)]. 
\bibitem{agg} V. V. Afonin, Y. M. Galperin, and V. L. Gurevich, 
Sov. Phys. JETP {\bf 61},1130 (1985)[Zh. Eksp. Teor. Fiz. {\bf 88} 
1906 (1985)]. 
\bibitem{agg1} V. V. Afonin, Y. M. Galperin, and V. L. Gurevich, \prb 
{\bf 40}, 2598 (1989). 
\bibitem{kr} V. I. Kozub and A. M. Rudin, \prb 
{\bf 55}, 259 (1997). 
\bibitem{Imry} Y. Imry, in: \emph{Tunneling in Solids, Proceedings of 
    the 1967 NATO Conference}. edited by E. Burstein and S. Lundquist 
  (Plenum Press, New York) ch. 35 (1969). 
\bibitem{Phillips1} W. A. Phillips, \prl {\bf 61}, 
2632 (1988). 
\bibitem {abric}A. A. Abrikosov, Physics {\bf 2} 21(1965). 
\bibitem{maleev} S. V. Maleev, Sov. Phys. JETP, {\bf 57}, 149 
(1983)[Zh. Eksp. Teor. Fiz. {\bf 84}, 260 (1983)]. 
\bibitem{maleev1} S. V. Maleev, Theor. and Math. Physics, {\bf ???}, 694 (1970) 
[Teor. Mat. Fiz. {\bf 4}, 86 (1970)]. 
\bibitem{evans} G. C. Evans, Rend. R./Accad. dei Lincei {\bf 20}, 453 (1911). 
\bibitem{Phillips} W. A. Phillips, J. Low Tepm. Phys. {\bf 7}, 351 (1972). 
\bibitem{Cornell1} K. S. Ralls and R. A. Buhrman, \prl 
{\bf 60}, 2434 (1988) 
\bibitem{Cornell2} D. C. Ralph and R. A. Buhrman, \prl 
{\bf 69}, 2118 (1992). 

\bibitem{Aleiner} I. L. Aleiner, B. L. Altshuler, Y .M. Galperin, \prb 
   {\bf 63}, 201401(R) (2001). 
\bibitem{kondo} I. L. Aleiner, B. L. Altshuler, Y. M. Galperin 
\prl  {\bf 86}, 2629 (2000). 

\bibitem{Polymer} A. N. Aleshin, V. I. Kozub, D.-S. Suh, Y. W. Park, 
  \prb 
 {\bf 64} (2001). 
\bibitem{Chandrasekhar} S. Chandrasekhar, Rev. \ Mod. \ Phys. {\bf
    15}, 1 (1943). 
\end{thebibliography}
\end{document}